\documentclass[12pt]{iopart}
\usepackage[english]{babel}
\usepackage{amssymb}
\usepackage{graphicx}
\usepackage{euscript}
\usepackage{indentfirst}
\usepackage{wrapfig}
\usepackage{latexsym}

\def\eps{\varepsilon}
\def\ffi{\varphi}
\def\x{{\mathbf x}}
\def\z{{\mathbf z}}
\def\q{{\mathbf q}}

\def\c{{\mathbf c}}

\def\V{{\mathbf V}}
\def\q{{\mathbf q}}

\def\ZZ{\EuScript{Z}}
\def\FF{\EuScript{F}}

\newcommand{\pd}{\partial}

\begin{document}
\title[Renormalization constants in
Kraichnan model]{Convergence of perturbation series for
renormalization constants in Kraichnan model with "frozen"
velocity field}
\author{M V Komarova$^1$, I S
Kremnev$^2$ and M Yu Nalimov$^3$}

\address{{}$^{1,3}$ St.
Petersburg State University, Physical Faculty, Ulyanovskaya
Street, 3, 198504, Petrodvorets, St. Petersburg,
Russia}
\address{{}$^2$Institute for Analytical
Instrumentation RAS, Rizhsky Prospect, 26, 198103, St.
Petersburg, Russia}
\eads{\mailto{$^1$ komarova1@paloma.spbu.ru},
\mailto{$^2$ ilya.kremnev@gmail.com},
\mailto{$^3$ Mikhail.Nalimov@pobox.spbu.ru}}

\begin{abstract}

Instanton was found for Kraichnan model with 'frozen' velocity field.
Large order asymptotic of quantum-field perturbation expansion
for renormalization constant $Z_\nu$  was investigated.  It was shown that
this expansion is convergent one. The radius of convergence
was calculated.
\end{abstract}

\section{Introduction}
Large-order asymptotic analysis of quantum-field
perturbation expansions is an actual problem of modern
statistical physics. Direct perturbation calculations are
cumbersome and difficult. Knowledge of large-order asymptotic and choice of  the corresponding
resummation procedure allow to get more or less good
estimation on the base of a few first terms of perturbation
series. Application of the resummation procedure without
knowledge about large-order asymptotic behaviour can lead
to inaccurate results.

Article \cite{Basuev} can be considered as a first attempt
to solve this problem. The asymptotes were
estimated here merely by number of graphs in the perturbation
expansion order. The correct large-order asymptotic investigation
proposed by Lipatov \cite{Lipatov} is based on the
saddle-point calculation of path integrals. It was used for
all main quantum-field theory models and static models of critical
behaviour, see \cite{Borel}.  Large-order analysis also was
consistently constructed for the dynamic models with
equilibrium static limit \cite{A-H, A-H1} using standard
Martin-Siggia-Rose (MSR) \cite{MSR} variables.  For all
models mentioned above the instanton was found and divergent
character of perturbation series was proved.  Then there is
common opinion that the instanton existence always leads to
divergent series.

This paper accounts the Kraichnan model with 'frozen' velocity field. It
describes the turbulent diffusion in stationary random field
\cite{Orzak Yakot,Juha1,Juha2} and well known problem of random walks in
random media \cite{Marinari1} -- \cite{Kravtsov2}.
The renormalization group (RG) approach was used for investigation of the scaling behavior in this model \cite{Juha1,Juha2}. The objects of interest are renormalization constants. In this paper large order
asymptotic of perturbation expansion of viscosity
renormalization constant in this model is investigated on the base of instanton found and it
is shown that the coefficients of perturbation series grow
essentially slower than $N!$. The series of perturbation
theory turn out to be convergent. The radius of convergence is
calculated.

Then our result contradicts the common opinion about
connection between instanton existence and
perturbation expansion divergence. The more or the
less general explanations of this phenomenon is given in Sec.
\ref{4}. Let us note that the examples of instanton analysis
of convergent series are known. One of these is the convergent
perturbation series introduced in \cite{Usveridze}. The instanton analysis
of the convergence of this expansion was fulfilled in \cite{Juha-U}.
Another example is standard Kraichnan model \cite{Obuhov, Kraichnan}.

The large order asymptotic analysis is more difficult in dynamic models then in static ones. Usually there is no instanton in natural class of MSR
variables here. For example the absence of instanton  in
Kraichnan model of turbulent diffusion was proven
in \cite{Balk}. Fortunately, the Lagrange variables can
be used in this model \cite{Chertkov, AndKomNal}. In
these variables instanton was found and the large-order
asymptotic behaviour was investigated \cite{AndKomNal}.
The perturbation series in this model appeared to have a
finite radius of convergence. Note that a specific feature of
standard Kraichnan model is a proportionality of the velocity field
correlator to $\delta(t)$-function.
Then a lot of graphs of perturbation theory are absent here
\cite{Adzhemyan} that could explain the convergence of series
discussed.

Because of the difficulties of instanton analysis the
large-order asymptotic form was estimated in some papers
merely by the number of graphs at large order of perturbation
expansion in spirit of \cite{Basuev}. This approach produces
accurate results for models with scalar and vector fields
without derivations in interaction. But  we will show that
such estimation \cite{Orzak Yakot} may cause a mistake.
Kraichnan model with frozen velocity
field discussed in this article is a good example of this
fact. In contrast with standard Kraichnan model, here
the velocity field correlator doesn't depend on time, so the number of
perturbation diagrams demonstrates a factorial behaviour $N!$
while the perturbation series is convergent. Simplified
example of  model considered with the constant velocity was
considered in \cite{TMPH}, where the number of
perturbation diagrams demonstrates a factorial behaviour too,
the instanton was found and the perturbation series
convergence was proved by exact solution of the model.

This paper is organized as follows. The Kraichnan model with a "frozen"
velocity field is described briefly in Sec. \ref{Model}. Also the response
function to be studied using MSR-formalism is introduced. The composite
operator in  Lagrange variables which is used for calculation of
renormalization constant $Z_\nu$ is introduced in Sec. \ref{Lagrange}.
Instanton approach for this composite operator is presented in Sec. \ref{4}.
A  particular solution of stationary equations used in the following analyses
is investigated in Sec. \ref{partsol}. The renormalization of the Green
function with the composite operator is described in Sec. \ref{simplepoles}.
Large-order asymptotic for expansion of $\ln Z_\nu$ is calculated in
Sec. \ref{replicatrick} using the replica trick. Cumbersome
stationarity equations for some numerical parameters are
discussed in Appendix.

\section{Kraichnan model with a 'frozen' velocity field}
\label{Model}

Kraichnan model describes the turbulent advection of passive scalar
admixture in $d$-dimensional fluid. It is based on a
stochastic equation
\begin{equation}
\label{kreycnan_eq}
[\partial_t+ g {\bf \nabla }{}_iV_i(\x, t)-\nu \Delta
]\varphi(\x,t)=\xi(\x,t).
\end{equation}
Here $\x\in \mathbb{R}^d$ and $t$ are  space and time
variables, $\ffi (\x,t)$ is a passive scalar field, $\V (\x, t)$
is a random vector velocity field, $\xi (\x,t)$ is a random
force, $\nu$ is a viscosity, $g$ is a
coupling constant. Laplacian and gradient refer to the space
variable $\x$;  here and henceforth all derivatives in squared
brackets act on the field $\varphi$ as well. For shortness we
introduce $\partial_t \equiv \partial/\partial t$. The
convolution with respect to repeating subscripts is implied
here and henceforth.

The random values $\xi$ and $\V$ are supposed to be distributed by Gauss low. It is known that the results obtained in RG analysis are independent from $D_\xi$ \cite{Juha1,Juha2} then one can state the correlator $D_\xi$ has an arbitrary form.  In contrast with
the standard Kraichnan model in the model discussed the velocity field correlator doesn't
depend on time
$$\langle\V_i(\x,t)\V_j(\x',t')\rangle
=D_{ij}(\x-\x')$$
(compare with
$\langle\V_i(\x,t)\V_j(\x',t')\rangle =D_{ij}(\x-\x')
\delta(t-t')$ in the standard Kraichnan model), in other words one can consider $\V$-field as a time independent.  The
velocity field correlator in the momentum representation has the
power-like form \cite{Juha1,Juha2}
\begin{equation}
\label{D(q)}
D_{ij}^F({\bf q}) \equiv\int d^d{\bf z}D_{ij}({\bf z})\exp(i{\bf qz})
=\lambda_T\Big(\delta_{ij}-\frac{\q_i\q_j}{\q^2}\Big)\frac{1}
{\q^{2\alpha}}+\lambda_L\frac{\q_i\q_j}{\q^2}\frac{1}{\q^{2\alpha}},
\end{equation}
that was used for RG-analyses of the model. Then in
the coordinate representation
\begin{eqnarray}
\label{D(x)1}
D_{ij}(\z)=a_1\frac{\delta_{ij}}{\z^{2\beta}}+a_2\frac{\z_i\z_j}
{\z^{2\beta+2}},\qquad \beta=d/2-\alpha=1-\eps/2,
\end{eqnarray}
where the parameters $a_1$, $a_2$ are known:
\begin{eqnarray}
\label{a1}
a_1=\frac{\Gamma (\beta)}{2^{2\alpha+1}\pi^{d/2}\Gamma(\alpha +
1)} (\lambda_T(2\alpha - 1) + \lambda_L),\\
\label{a2}
a_2=(\lambda_T - \lambda_L) \frac{\Gamma(\beta+1)}
{2^{2\alpha}\pi^{d/2}\Gamma(\alpha+1)},
\end{eqnarray}
$\lambda_T$ and $\lambda_L$ are transverse and longitudinal
coupling constants. The parameter $\lambda_L$ corresponds to compressibility
of fluid.

The model considered is important for the description of
 diffusion in random fluids \cite{Juha1,Juha2}. Moreover it
relates to the developed turbulence problem \cite{Bouch}. In fact the model (\ref{kreycnan_eq}, \ref{D(q)}) has two coupling constants $\lambda_T$, $\lambda_L$. Nevertheless for large order asymptote of $\eps $
expansion investigation these can be reduced to the only one
coupling constant $g$. Indeed in a fixed point
all coupling constants are proportional to a small parameter
of regular expansion ${\eps=2+2\alpha-d}$, \cite{Juha1,
Juha2} $\lambda_T\sim\lambda_L\sim g^2$.  The similar
situation was observed in dynamic models C-H (A, B, C,... are a common
nomination for particular dynamic models introduced   in
\cite{Halperin}) with equilibrium $\ffi^4$ static limit
\cite{Vasiljev, A-H1}.

The infrared behaviour of the model was investigated by
means of RG method. The renormalization yields the substitution
$\nu\to \nu Z_\nu $, $g\to gZ_g$,
the renormalization constant $Z_\nu$ is calculated by means of
perturbation theory, then it has a form of series in
coupling constant. The properties of this series
are the main subject of this paper.

Usually MSR-formalism \cite{MSR, Vasiljev} is used to transform
stochastic models to quantum-field ones. The base equation
(\ref{kreycnan_eq}) is represented in a form of path integration
in auxiliary field $\ffi'$; path integrations in $\ffi$ and
$\V$ fields are introduced to study averaged characteristics
of fluid. Then the expression for the response function in an
arbitrary velocity field has a form
\begin{eqnarray}
\label{Gv}
G_V^{(1,2)}= \frac  {\int \EuScript{D} \varphi \EuScript{D} \varphi '
\hskip 0.2cm
\varphi(\x_1,t_1)\varphi'(\x_2,t_2)\exp(S^{{\hskip0.3mm}msr})}
{\int \EuScript{D} \varphi \EuScript{D} \varphi '
\exp(S\lefteqn{{}^{{\hskip1mm}msr}}|_{g=0})},
\end{eqnarray}
with the renormalized action \cite{Juha1, Juha2}
\begin{eqnarray}
\label{MSR}
S^{{\hskip0.3mm}msr}= \frac{ \varphi'(\x,t)D_\xi\varphi'(\x',t')} {2}
+ \varphi'(\x,t)[\partial_t +gZ_g{\bf \nabla}_i\V_i(\x) -
\nu Z_\nu \Delta] \varphi(\x,t),
\end{eqnarray}
usual standard agreements for dynamic models
\cite{Vasiljev} and all integrations needed are implied
henceforth. The normalization factor in Exp. (\ref{Gv})
corresponds to a free model (at $g=0$).

After the integration in $\V$ field one obtains MSR
representation for the renormalized response
function
\begin{eqnarray}
\label{foop}
<\ffi(\x_1,t_1)\ffi'(\x_2,t_2)>= \frac
{\int \EuScript{D}\V G_V^{(1,2)}
e^{-\V_i(\x)D_{ij}^{-1}(\x-\x')\V_ j (\x') /2} }
{\int \EuScript{D}\V e^{-\V_i(\x)D_{ij}^{-1}(\x-\x')\V_j (\x')/ 2}}.
\end{eqnarray}

\section{Lagrange variables}
\label{Lagrange}

 As it was stated in papers \cite{AndKomNal, Balk}
 there is no instanton for Kraichnan model in the framework of MSR formalism.
Similar arguments are correct for the model with the frozen
velocity field. But Lagrange variables \cite{Chertkov,
AndKomNal}  can be used for instanton analysis of the model
considered as in standard Kraichnan model.

Seems these  variables have an origin in the quantum-fields
methods application in random walks and macromolecules
problems (see \cite{Freed}) in analogy with Hamiltonian form of
the standard Feynman-Kac path integral.
 Lagrange variables can be introduced
successfully in the dynamic models with linear in main field
$\ffi$ stochastic equation (\ref{kreycnan_eq}) only. Then
the dynamic equation for the response function (\ref{Gv}) can
be considered as  Schrodinger equation \cite{Chertkov} or as
Fokker-Plank  equation \cite{AndKomNal}, and response
function in an arbitrary velocity field can be represented in
the form
\begin{equation}
\label{GVLag}
G_V^{(1,2)}=\frac{\Theta(t_2-t_1)}{(4\pi \nu)^{d/2} (t_2 -t_1)^{d/2}}\frac
{\int \EuScript{D} \c \EuScript{D} \c '
\hskip 0.2cm \exp(S^{{\hskip0.3mm}Lgr})}
{\int \EuScript{D} \c \EuScript{D} \c '
\hskip 0.2cm
\exp(S\lefteqn{{}^{{\hskip1mm}Lgr}}|_{g=0})},
\end{equation}
$$S^{Lgr}=\int\limits_{t_1}^{t_2} d\tau\Bigg(
-\nu Z_\nu \c'^2(\tau)+i\c'(\tau )\partial_\tau \c(\tau)+gZ_g \c'(\tau)
\V(\c(\tau) )\Bigg).$$
The  boundary conditions
$$ \c(t_1)=\x_1,\quad \c(t_2)=\x_2$$
are implied for the path integration in $\c$ field in the numerator.
The integration in $\c'$ fields is supposed to have free boundary
conditions. The normalization path integral  corresponds to a free
model with $g=0$ and may be calculated at zero boundary
conditions for $\c$, $\c'$ fields. Vector fields $\c(\tau)$, $\c'(\tau)$ play a role of
coordinates and momenta of fluid particles and depend on time
only.

Let us note that this representation produces one Green
function of the model only, namely the response Green function
in an arbitrary velocity field. Then the MSR action (\ref{MSR})
can  be obtained from (\ref{GVLag}) by no change of
variables and the statement about an instanton absence in MSR
variables is not correct for Lagrange ones.

Lagrange variables can be used to investigate
renormalization constant $Z_\nu$ in theory (\ref{MSR}). Indeed
let's differentiate Exp.~(\ref{foop}) with respect to $\nu$ in
order to extract $Z_\nu$ constant.  Then one has two point
Green function with the   $\ffi'\nu Z_\nu \Delta\ffi$
composite operator insertion. Using the response function
(\ref{foop}) this Green function
$$Z_\nu <\ffi(\x_1,t_1)\ffi'(\x_2,t_2)
\int d\x_0 dt_0\ffi'(\x_0,t_0)\Delta \ffi(\x_0,t_0)>\equiv
 G$$
can be rewritten as
\begin{eqnarray}
\label{www}
G=Z_\nu \int d\x_0 dt_0\\
\nonumber
\frac {\int
\EuScript{D}\V G_V^{(1,2)}(\x_1,t_1,\x_0,t_0)\Delta G_V^{(1,2)}
(\x_0,t_0,\x_2,t_2)
e^{-\V_i(\x)D_{ij}^{-1}(\x-\x')\V_ j (\x') /2} }
{\int \EuScript{D}\V e^{-\V_i(\x)D_{ij}^{-1}(\x-\x')\V_j (\x')/ 2}}.
\end{eqnarray}

The renormalization  constant $Z_\nu$ has poles  in $\eps$ at
${\eps\to 0}$ that have to cancel UV divergences of the model.
As we consider a renormalized response function (\ref{foop})
Exp.~(\ref{www}) must be finite at $\eps\to 0$. Then
\begin{equation}
\label{inform}
\mathop{\mbox{res}}\limits_{\varepsilon\to 0}\ln Z_{\nu}=
-\mathop{\mbox{res}}\limits_{\varepsilon\to 0}\ln \int d\x_0 dt_0 G
\end{equation}
and the Green function $G$ contains all information needed
about the poles of renormalization constant $Z_{\nu}$.

The diagrams for $G$ include the internal loop part and the external full
propagator part without divergences. The last  does not contribute
into Exp. (\ref{inform}) then one will discuss now the amputated
diagrams for $G$.

The Green function $G$ can be easily written in Lagrange
variables. The following Gaussian path integration in field~$\V$ and
Fourier transforms in $\x_2-\x_1$ and $t_2-t_1$ variables yield
the action to be studied
\begin{equation}
\label{S}
S=-i{\bf q}(\x_2-\x_1)-\nu Z_{\nu}(\c_1'^2+\c_2'^2)+ i\c_1'\partial
\c_1+i\c_2'\partial \c_2+Z_uS_u,
\end{equation}
where $\bf q$ is a momentum. Frequency $\omega =0$, that is sufficient
for the renormalization constant investigation, because this
constant is frequency independent. A nonlinear
part of the action is collected in the term
$$S_u=-\frac{u}{2}\Bigg(
c_{1i}'(\tau_1)D_{ij}(\c_1(\tau_1)-\c_1(\tau_1'))c_{1j}'(\tau_1')+$$
$$+ c_{2i}'(\tau_2)D_{ij}(\c_2(\tau_2)-\c_2(\tau_2'))c_{2j}'(\tau_2')+
2c_{1i}'D_{ij}(\c_1-\c_2)c_{2j}'\Bigg),\qquad u\equiv g^2.$$
Here and henceforth one implies that the fields $\c_l$, $\c_l'$
($l=1,2$) with argument omitted depend on $\tau_l$ . All
necessary integrations in $\tau_l$ and the ranges of
integration
$$t_1\le \tau_1,\tau_1'\le t_0 \le \tau_2,\tau_2'\le t_2$$
are assumed. Finally one gets for Fourier transformed $G$
function the following expression in Lagrange variables:
\begin{eqnarray}
\label{G-Lagr}
G=\int d(\x_2-\x_1)\int d(t_2-t_1)\frac{1}{\aleph(4\pi\nu
\sqrt{(t_2-t_0)(t_0-t_1)})^d}\\
\nonumber
\int\limits^{\c_1(t_1)=\x_1,
\hskip0.1cm \c_2(t_2)=\x_2,} _{\c_1(t_0)=\c_2(t_0)=\x_0}
\EuScript{D} \c_1\EuScript{D} \c_2 \EuScript{D} \c_1
'\EuScript{D} \c_2 ' \hskip 0.2cm W  e^{S},
\end{eqnarray}
where fore-exponential factor
$$W =-\frac{\int d\tau_1 d\tau_2(i\c_1'+(t_1-\tau_2)\FF_1)
\int d\tau_1 d\tau_2(i\c_2'-(t_2-\tau_2)\FF_1)}{(t_1-t_0)(t_2-t_0)}+$$
\begin{equation}
\label{W}
+\frac{\int d\tau_1 d\tau_2(t_1-\tau_1)(t_2-\tau_2)\FF_2}
{(t_1-t_0)(t_2-t_0)},
\end{equation}
$$
\FF_s\equiv u Z_u\c_{1i}'
\frac{\partial^s}{\partial\c_1^s}D_{ij}(\c_1-\c_2)\c_{2j}
$$
is produced by Laplace operator in (\ref{www}).
Let us note that  the non-trivial boundary conditions for $\c_1$, $\c_2$
fields
\begin{equation}
\label{bound}
{\c_1(t_1)=\x_1, \qquad \c_2(t_2)=\x_2,}\qquad
{\c_1(t_0)=\c_2(t_0)=\x_0}
\end{equation}
mean the path integration with fixed boundary conditions.
The integration in $\c_l'$ fields is supposed to have free boundary
conditions. The normalization factor $\aleph$
corresponds to a free model with $u=0$ and may be calculated at
zero boundary conditions for $\c_l$:
\begin{equation}
\label{ALEPH}
\aleph=
\int\limits^{\c_1(t_1)=\c_2(t_2)=0}_{\c_1(t_0)=\c_2(t_0)=0}
\EuScript{D} \c_1\EuScript{D} \c_2 \EuScript{D} \c_1
'\EuScript{D} \c_2 ' \hskip 0.2cm \exp{(S|_{g=0})}.
\end{equation}

For shortness we introduce the parameters
\begin{equation}
\label{1}
T\equiv t_2-t_1,\quad T_1\equiv t_0-t_1,\quad T_2\equiv
t_2-t_0,
\end{equation}
\begin{equation}
\label{2}
\x\equiv \x_2-\x_1,\quad \x^{(1)}\equiv \x_0-\x_1,\quad
\x^{(2)}\equiv \x_2-\x_0,
\end{equation}
the values $x$,  $x_0$, $x_1$, $x_2$, $x^{(1)}$, $x^{(2)}$
are the modules of corresponding vectors.

\section{Instanton analysis}
\label{4}

In order to extract the $N$-th term of perturbation series
the Cauchy formula is traditionally used
\cite{Lipatov}:
\begin{equation}
\label{cashi}
G^{[N]}= \frac{1}{2\pi i} \oint \frac{G(u)}{u^{N+1}} du,
\qquad
G(u)=\sum\limits_{N=0}^\infty G^{[N]}u^N,
\end{equation}
the integration is produced along a closed contour containing
zero in a complex plane.

Let us extract the large $N$ parameter from the action~$S$
(\ref{S}) by the following scaling
\begin{equation}
\label{tjazh}
\{\c_l,\c'_l\}\to \{N^{1/2}\c_l, N^{1/2}\c_l'\},    \qquad
\x  \to  \sqrt{N}\x , \qquad   u \to  N^{\beta} u,
\end{equation}
the scaling of $\x$ variable is necessary due to the
connection between $\c$ and $\x$ based on (\ref{bound}).
The same scaling in normalization factor $\aleph$ cancels the
determinant corresponding to this change of variables. The
momentum $\q$ is scaled $\q \to N^{1/2}\q$ also. It is possible
because the renormalization constant investigated is momentum
independent.

The integrals in $\c'$, $\c$ and $u$ at large $N$ can be calculated by
the saddle-point approach. The main contribution at $N\to
\infty$ is given by the integration near the {\it {instanton}}
that is a special realization of variables $\c_{st}$,
$\c'_{st}$, $u_{st}$. The action has an extremum at the
instanton.

Let us mark that the scaling procedure is an essential in the
determination of the connection between instanton existence
and divergence or convergence of perturbation theory.
Usually the quantum-field theory action has a form
$S=S_0+S_{int}$, where $S_0$ is a free part of the action,
$S_0=\Phi K \Phi /2$ where $\Phi$ denotes a field or a set of
fields of the theory considered, $K$ is a linear operator.
$S_{int}$ is an interaction of the form $S_{int}=\lambda\Phi ^k$
($k>2$) in a local theory. Presence of derivations in the
interaction does not affect analysis presented below.
$\lambda$ is a coupling constant and the expansion parameter.
To extract large $N$ parameter from the action the
scaling $\Phi \to \sqrt{N}\Phi$, $\lambda \to \lambda
/N^{(k-2)/2}$ is necessary. Then the main exponential
contribution in expression similar to (\ref{cashi}) due to
instanton is proportional to
$$ G^{[N]}\sim N^{\frac {k-2}2N}e^{-NS_{st}},
$$
where $S_{st}$ is an action in the stationary point. This
contribution demonstrates the divergence of the perturbation
expansion in $\lambda$ due to $k>2$. Then the divergence of
the series is connected with the scaling of the coupling
constant. In \cite{Borel} another instanton analysis scheme
without Cauchy formula using was proposed. But the results of
analysis presented above will be the same in both schemes.

Note that  the model considered (\ref{S}) in Lagrangian
variables differs from the general case. The coupling constant
is scaled here by positive power of $N$ due to the nonlocal
character of the interaction, then instanton analysis can
lead to convergent series as in standard Kraichnan model
\cite{AndKomNal} or simplified example of Kraichnan model with
constant velocity field \cite{TMPH}, where the instanton was found and
the perturbation series convergence was proved by exact
solution of the model.

Let us describe instanton calculation in the model (\ref{S}) in
more details. It's quite reasonable to simplify problem by
taking into account the symmetry of the model that is
initially violated by $\x$ vector only.  Then we suppose that
the fields $\c_l$, $\c'_l$ are parallel to $\x$ and the only
modules $c_l$, $c_l'$ must be found.  Let us mark that the
stationarity equations are non-linear differential ones. The
existence and the uniqueness of the solution are not proved in
general case.  We propose to use the solution with the same
symmetry as Green function investigated. Note that the
spherical symmetry of the instanton in Lipatov work
\cite{Lipatov} based on the same ideas.  The possibility of
other solutions existence with other contributions to large
order asymptotes is an open question.  The same situation is
observed in every case of instanton analysis.
The supposition about the symmetry of solution used here was proven in
\cite{TMPH} for simplified Kraichnan model with known exact solution.
For standard Kraichnan model considered in \cite{AndKomNal}
this supposition yields the result coinciding with the exact known
anomalous dimensions of a set of composite operators.

All stationarity equations are supposed then to be projected on
the direction of $\x$ vector.
It also simplifies the tensor structure of $D$ correlator
that has more compact form now:
\begin{equation}
\label{D(x)}
D(\x) = \frac{D_0}{|\x|^{2\beta}}, \qquad
D_0=a_1 + a_2.
\end{equation}

Except the regular in $\eps$ terms the action $S$ (\ref{S})
contains poles in $\eps$ due to  $Z_\nu$, $Z_g$ constants.
It was  shown in \cite{Our1} that while a renormalization constant is investigated  the
corresponding singularities must be extracted before any
instanton calculations and later they contribute only in a
fore-exponential factor of the saddle-point method.  This
extraction of singularities is necessitated by existence of
two large parameters, namely $1/\eps$ connected with
regularization and the saddle-point method parameter $N$.
Due to the renormalization approach the value $N\eps$ must be
considered as a small one in the framework of instanton
analyses \cite{Our1}.  Then the exponential term in
(\ref{G-Lagr}) must be presented in a form
\begin{equation}
\label{series}
\exp(S)=\exp(S_{reg}+S_{sing})=\exp(S_{reg})\sum\limits_{p=0}^{\infty}
\frac{1}{p!}(S_{sing})^p,
\end{equation}
$$S_{reg}=S\Big|_{\hskip-0.2cm
\begin{array}{c}\mathop{}\limits_{Z_\nu =1}^{Z_u=1}\end{array}},\qquad
S_{sing}=-\nu (Z_\nu-1)(\c_1'^2+\c_2'^2)+(Z_u-1)S_u$$
and only the term $S_{reg}$ in the exponent must be variated. In standard
Kraichnan model this approach was proven in \cite{AndKomNal} by the
comparison of the radius of convergence calculated with the exact
known results. Mention should be made that the l.h.s and the r.h.s
of the identity (\ref{series}) are essentially different
under the integral $\oint du/u^{N+1}$ and they yield different results
of the saddle-point method due to the
competing of parameters $N$ and $1/\eps$.

Let's remind the renormalization constants $Z_\nu$, $Z_u$ in minimal subtraction (MS) scheme have a form $1 + $ poles in $\eps$ terms, so that $(Z-1)$ contain the pure singularities at $\eps\to 0$ only. Exp.~(\ref{series}) shows that the divergences in $\eps$ make a sense in the framework of perturbation theory and diagrammatic expansion only. Then the path integration must be interpreted as a sum of perturbation terms, the renormalization is supposed to be a cancellation of divergences.

Then let us set $Z_\nu = 1, Z_g = 1$ in the expression for the
action~(\ref{S}) in order to write the regular instanton equations.
The variations of action $S$ involve the integral operators of the form
$$[D_{lk}\c_k'](\zeta) \equiv\int d\tau_k D(\c_l(\zeta)-\c_k)c_k',\qquad
l,k=1,2.$$
For example the variation in $\c_1$ yields
\begin{equation}
\label{24}
\frac{\delta S}{\delta \c_1(\zeta)}=0\qquad\Rightarrow\qquad
{uc_1'(\zeta)}\partial_{\zeta}\left([D_{11}\c_1'](\zeta)+[D_{12}\c_2']
(\zeta)\right)=-{i\partial_{\zeta}c_1'\partial_{\zeta}c_1},
\end{equation}
as well the variation in $\c_1'$
\begin{equation}
\label{23}
\frac{\delta S}{\delta \c_1'(\zeta)}=0\qquad \Rightarrow \qquad
-2\nu c_1'(\zeta) +i\partial_{\zeta}c_1
- u([D_{11}\c_1'](\zeta)+[D_{12}\c_2'](\zeta))=0.
\end{equation}

The contribution of integral operators in eq.~(\ref{24}) can
be excluded with the help of eq.~(\ref{23}). For this purpose
we should differentiate~(\ref{24}) in $\zeta$
$$\partial_{\zeta}\frac{\delta S}{\delta \c_1'(\zeta)}=0
\quad \Rightarrow\quad
-2\nu \partial_{\zeta}c_1'+i\partial^2_{\zeta}c_1-
u\partial_{\zeta}\left([D_{11}\c_1']+[D_{12}\c_2']\right)=0,
$$
express $[D_{11}\c_1']+[D_{12}\c_2']$ and substitute it in
the eq.~(\ref{24}). It yields the 2nd order differential
instanton equation
\begin{equation}
\label{2ndorder}
i\partial^2_{\zeta}c_1-2\nu \partial_{\zeta}c_1'+
i\frac{\partial_{\zeta}c_1'\partial_{\zeta}c_1}{c_1'(\zeta)}=0
\end{equation}
that can be solved with respect to $\pd_{\zeta}c(\zeta)$. The same
calculation can be done for $c_2$ field. The solution of
equation (\ref{2ndorder}) has a form
$$\pd_{\zeta}c_l(\zeta)=\frac{F}{c'_l(\zeta)}-i\nu
c_l'(\zeta),\qquad l=1,2$$
and contains the arbitrary parameter $F$. Each numeric value of $F$ constant
corresponds to a particular solution with its own boundary
conditions.

The following calculations at arbitrary $F$ can
not be performed analytically,  the simplest case at
$F = 0$ allows to reduce the instanton to quadrature only.
Nevertheless the corresponding particular solution can be used
to determine the asymptotic behaviour of the renormalization
constant investigated.

\section{Particular solution}
\label{partsol}
Let us consider the particular solution with $F=0$ and determine
its boundary conditions. The solution
\begin{equation}
\label{sol}
c_l'(\zeta)=\frac{i\pd_\zeta c_l}{\nu },\qquad l=1,2
\end{equation}
can be substituted into the variation equations with respect to $\c_l'$.
Using the identity $d\tau_l\pd c_l=dc_l$ the result can be written
in a form
$$-\pd_{\zeta}c_l=\frac{u}{\nu}\int\limits_{x_1}^{x_2}
D(c_l(\zeta)-z)dz,\qquad l=1,2.$$
The last differential equation can be easily integrated and this
leads to the solution for the $c_l(\zeta)$ fields in quadrature:
\begin{equation}
\label{fbound}
\int\limits_{x_1}^{c_1(\zeta)}\frac{dc}{\int_{x_1}^{x_2}D(c-z)dz}=
-\frac{u(\zeta-t_1)}{\nu},\quad
\int\limits_{x_0}^{c_2(\zeta)}\frac{d c}{\int_{x_1}^{x_2}D(c-z)dz}=
-\frac{u(\zeta-t_0)}{\nu}.
\end{equation}
Note that an analytic regularization is assumed in (\ref{fbound}) that eliminates the
singularity in $z=c$ point.

After the substitution of explicit form $D(\c)$ (\ref{D(x)})
Exp. (\ref{fbound}) produces the boundary condition
interested
\begin{equation}
\label{bound1}
\frac{1}{T_1}\int\limits_0^{x^{(1)}/x}f(v)dv=\frac{uD_0}{x^{2-\eps}
(1-\eps)\nu}=\frac{1}{T_2}\int\limits_0^{x^{(2)}/x}f(v)dv,
\end{equation}
where the function $f(v)$ introduced is
$$f(v)=\frac{v^{1-\eps}(1-v)^{1-\eps}}{v^{1-\eps}+(1-v)^{1-\eps}}.$$

One sees that the case $F=0$ is simple enough to give a quadrature
representation (\ref{fbound}) for $\c_1$, $\c_2$ fields with the
boundary conditions (\ref{bound}).

Initially the problem consists in calculation of instanton for an arbitrary
boundary condition $\x(T)$. Exp. (\ref{bound1}) solves the
problem in the specific case with a special value of $
x^{(1)}$, $x^{(2)}$, and $x$. Then one has instanton for functional
integral in $\bf c$,
$\bf c'$ fields. But the object investigated (\ref{cashi},
\ref{S}) includes integrations in $\x$, $T$, $\x_0$, $t_0$
and $u$ also.
Our main idea here is to explore the independence of the renormalization
constant investigated on the momentum $\q$. Let us include the
integrals in variables $\x$, $\x_0$, $t_0$ into the
saddle-point method. Let's choose the momentum $\q$ so that
the solution of a stationarity equation for $\x$ variable be
exactly equal to the result obtained with the help of boundary
condition (\ref{bound1}) corresponding to the case $F=0$. This
choice gives us a chance to explore the particular solution
constructed analytically and to solve the problem without
numerical calculations. Moreover constant $F$ is not a free
parameter in this approach, then one has no problem of zero
modes connected with $F$ arbitrariness.

The stationarity equations for $\x$,
$\x_0$, $t_0$ variables are too cumbersome to be written down here. 
Nevertheless, as shown in  Appendix of the article, these can be 
simplified  significantly using particular solution (\ref{sol}), 
its properties, and integration by parts.
Then the equations solution calculated
corresponds to the case $x^{(1)}=x^{(2)}$, $T_1=T_2$ and the
action has the following form in stationary point (\ref{sol})
\begin{equation}
\label{eq:S_st_c_st}
S_{st} = -i\q\x - \frac{uD_0}{\nu^2\eps(1-\eps)}x^\eps.
\end{equation}
Now the stationary equations for $\x$ in $F=0$ case:
\begin{equation}
\label{eq:dS/dx}
\frac{\delta S}{\delta \x}=0\quad\Rightarrow
\quad \frac{iq\nu^2 }{u}=
\int\limits_{0}^{x}D(z)dz.
\end{equation}
Besides we have the boundary conditions (\ref{bound1}) imposed by our
choice of particular solution $F=0$. By solving equation (\ref{eq:dS/dx})
one obtains the proper value for $q$ in case $F=0$:
\begin{equation}
\label{x_st}
q=q_0=\frac{iD_0u}{(1-\eps)x^{1-\eps}\nu^2 },\qquad
x^{2-\eps}=x_{st}^{2-\eps}=\frac{uD_0 T/2}
{(1-\eps)\nu \int\limits_0^{1/2}f(v)dv}.
\end{equation}
Combining (\ref{eq:S_st_c_st}) and (\ref{x_st}) finally, one obtains
the action $S$ (\ref{S}) at the stationarity solution
and $Z_\nu=1$, $Z_u=1$:
\begin{equation}
\label{Sst}
S_{st}= -\frac{u D_0 x_{st}^\eps}{\nu^2\eps}.
\end{equation}

\section{Simple poles in $\eps$}
\label{simplepoles}

Due to (\ref{inform})  simple poles in $\eps$ of $G$
contain all the necessary information
about the poles of renormalization constant $Z_\nu$ and the
corresponding critical indices.

Since $D_0(\eps)=a_1+a_2$ (\ref{D(x)}) and (\ref{a1},\ref{a2})
$D_0(\eps)$ function can be presented in a form
\begin{equation}
\label{AB}
D_0(\eps)\equiv A+\eps B(\eps),\quad B(\eps) = B_0 + B_1 \eps + O(\eps^2).
\end{equation}
It now follows that the action (\ref{Sst}) has a form
\begin{equation}
\label{ST}
S_{st}=-\frac{u A }{\nu^2\eps}
-\frac{u A (x_{st}^\eps-1)}{\nu^2\eps}
-\frac{u B(\eps) x_{st}^\eps}{\nu^2}.
\end{equation}

The first term here is singular in $\eps\to 0$, then
as well as $S_{sing}$ term discussed in Section~\ref{4} it
must be presented in fore-exponent form (\ref{series}).
The second term seems to be finite at small $\eps$. Nevertheless
its logarithmic behaviour in $x_{st}$ results in singular in
$\eps$ contribution to the large order asymptote. We
will discuss this at the end of this Section.
As a result the only regular term of action $S_{st}$ (\ref{ST})
is the third one. Then Exp. (\ref{series}) must be corrected by changing
$S_{reg}\to\bar{S}_{reg}$, $S_{sing}\to\bar{S}_{sing}$,
$$\bar{S}_{sing}=S_{sing}-\frac{u A }{\nu^2\eps}
-\frac{u A (x_{st}^\eps-1)}{\nu^2\eps}.$$
Combining  Exp. (\ref{ST}) with the  result for $x_{st}$ (\ref{x_st}) we get
$$\bar{S}_{reg}=(uT^{\eps/2})^{{2}/(2-\eps)} P(\eps),
\qquad P(\eps) \equiv -\frac{B(\eps)}{\nu^2}  \Bigg(\frac{D_0}
{(1-\eps)\nu 2\int_0^{1/2}f(v)dv}\Bigg)^{\eps/(2-\eps)}.$$
Thus the Green function  studied is of the form
$$G^{[N]}=N^{-N(1-\eps)}\oint \frac{du}{u^{N+1}}\int dT
 \ZZ(T,u,\eps)  \exp(NS_{reg}(\eps, T))\times$$
\begin{equation}
\label{simple}
\times \sum\limits_{p=0}^{\infty}\frac{1}{p!}
(\bar{S}_{sing})^p(1+O(N^{-1})).
\end{equation}

The $\ZZ$ factor stays for a Gaussian fluctuations
contribution and fore-exponential factor $W$ (\ref{W}), the corresponding
integration is normalized by $\aleph $ factor.
$O(N^{-1})$ term shows the accuracy of calculation
at $N\to \infty$.

The amputated Green function with composite operator considered must be
dimensionless. The factor $T^{-1}$ restoring this zero
dimension has to be produced by $\ZZ$ and we will extract
$T^{-1}$ from $\ZZ$ in order to stress this fact.

The integration in $T$ then diverges as a logarithm and produces
singularities. It can not be treated by the saddle-point
approach. The analogous situation exists in a well-developed instanton analyses for static $\ffi^4$ model where the role
of divergent integration parameter $T$ plays the scale
parameter in the coordinate space \cite{Lipatov,Our1}.

Let us change variables
$$\qquad \bar{u} \equiv uT^{\eps/2},$$
the new variable $\bar{u}$ is dimensionless. As a result the
expression
\begin{eqnarray}
\label{simple1}
\nonumber
G^{[N]}=N^{-N(1-\eps)} \int\frac{dT}{T^{1-N\eps/2}} \oint
\frac{d\bar{u}}{\bar{u}} \ZZ(\bar{u},\eps) \exp \Big{(}N\Big{[}
P(\eps) \bar{u}^{\frac{2}{2-\eps}} - \ln \bar{u} \Big{]} \Big{)}\times\\
\times \sum\limits_{p=0}^{\infty}\frac{1}{p!}
(\bar{S}_{sing})^p(1+O(N^{-1}))
\end{eqnarray}
can be integrated over $T$
in UV region (small $T$). This yields a simple pole
$2/(N\eps)$. The convergence of integral at large $T$ is
provided by IR regularization assumed. Note the factor $\ZZ$
does not contribute to the stationary equations as it does
not depend on~$N$. In the same way this factor did not affect
the simple pole in $N\eps$ at least at principal order in
$1/N$.

The integration in $\bar{u}$ is investigated by the
saddle-point approach. The stationary equation with respect
to $\bar{u}$ is
$$\frac{\partial \Big{[} P(\eps)
\bar{u}^{\frac{2}{2-\eps}} - \ln \bar{u} \Big{]}}{\partial
\bar{u}} = 0,\qquad \mbox{then} \qquad\bar{u}_{st}^{\frac{2}{2-\eps}}
= \frac{1-\eps/2}{P(\eps)}.
$$
So we obtain the leading order in $N$
$$G^{[N]} =C(\eps) N^{-N(1-\eps)+\rho} \frac{2}{N\eps}
e^{(\eps/2-1)}\Big{(} \frac{P(\eps)e}{1-\eps/2} \Big{)}^{(N+1)(1-\eps/2)}
\sum\limits_{p=0}^{\infty}\frac{1}{p!}(\bar{S}_{sing})^p.$$
The factor $C(\eps) N^\rho$ appears due to
$\ZZ(\bar{u}_{st},\eps)$ contribution
and the fluctuation integration in $\bar{u}$;
$\rho$ is a constant.

Let us discuss the residue in $\eps$ calculation.
Simple poles in $\eps$ can appear if higher poles of $\bar{S}_{sing}$
in the fore-exponent are multiplied by regular in $N\eps$ contribution
of the exponent term $\exp(S_{reg}(\eps))$.
Therefore in the MS scheme chosen all $p \neq 0$ terms of the sum contribute to the result. Fortunately a finite
renormalization could help us to map out all this terms.
Indeed let's scale the expansion parameter
$$u D_0(\eps) \rightarrow \varkappa(\eps) u D_0(\eps).$$
Such a renormalization is equivalent to a scaling of the velocity field
correlator and doesn't affect the scaling dimensions. Let's
choose $\varkappa(\eps)$ so that the regular part of action
loses its dependence on $N\eps$,
namely
$$\Big{(}\frac{P(\eps)e}{1-\eps/2}\Big{)}^{(1-\eps/2)}
\to \Big{(}\frac{P(\eps)e}{1-\eps/2}\Big{)}^{(1-\eps/2)}\Bigg|_{\eps=0}
\equiv K_0.
$$
The corresponding renormalization can be easily written as a perturbation
series
$$\varkappa(\eps) = 1 + \Big{(} \frac{B_0}{2A} \ln \Big{[}-\frac{B_0}{6A\nu}
\Big{]} - \frac{B_1}{A} \Big{)} \eps^2 + O(\eps^3),$$
the parameters $A$, $B_0$, $B_1$ are introduced in (\ref{AB}).
After the scaling all poles in $\eps$ contained in the sum
with respect to $p$ don't contribute to the simple pole.
As a result the residue discussed demonstrates the asymptotic
behaviour
$$\mathop{\mbox{res}}\limits_{\varepsilon\to
0}G^{[N]}=\mbox{Const}N^{Const}K_0^N/N!,\qquad
N\to\infty$$
that corresponds to a finite radius of convergence for
the perturbation series of $G$ function.

Now let's show that the second term in (\ref{ST}) doesn't change our answer.
Indeed  its behaviour in $x$ is logarithmic. Then the presence of additional
factor $\ln x\sim \ln(T)$ in (\ref{simple1}) is equivalent to the additional
operation $d/d(N\eps)$ of $T^{N\eps/2}$ factor in (\ref{simple1}).
This operation can't produce simple pole in $\eps$ as our
expression doesn't contain logarithmic in $\eps$ contributions. It was
shown in details for the similar problem in \cite{Our1}.

The following step is to calculate $\ln G(u)$, extract
the residue in $\eps=0$ of the $N$-th order of perturbation
theory and explore formula (\ref{inform}).  It's useful to
transform the logarithm of the composite operator $G$ with the help of replica trick.

\section{The replica trick}
\label{replicatrick}
The simplest way to present a logarithm of arbitrary integral
expression in an integral form is to use the formula
\begin{equation}
\label{zz}
\ln \int dT f(T)=\lim\limits _{r\to 0}\frac{\partial }{\partial
r}\prod\limits_{\alpha=0}^{r-1}\int dT_\alpha
f(T_\alpha).
\end{equation}
As a result the variable $T$ becomes an $r$-dimensional vector
in a replica space \cite{Vasiljev}.

Performing such a procedure with respect to $G$ one gets the
expression similar to (\ref{simple}) where the integrals must
be rewritten as follows
\begin{equation}
\label{repl}
\oint \frac{du}{u^{N+1}}\int(\prod\limits_{\alpha=0}^{r-1} dT_\alpha
)\exp(N\bar{S}_{reg}) \ZZ(\{T_\alpha\}_{\alpha=0}^{r-1}),
\end{equation}
$$\bar{S}_{reg}=\sum\limits_{\alpha=0}^{r-1}
(uT_\alpha^{\eps/2})^{2/(2-\eps)}P(\eps),$$
the factor $\ZZ$ depends on the replica variables
$\{T_\alpha\}$. Nevertheless one knows that the
expression investigated with the help of replica trick is
also dimensionless due to $\ZZ$ contribution. Besides the
expression must be proportional to $r$ at small $r$. Then the
operation $\lim\limits _{r\to 0}{\partial }/{\partial r}$ in
(\ref{zz}) yields a non-trivial finite result.

It is easy to see that the saddle-point approach can not be
applied to all integrations in (\ref{repl}).
At least one of integration considered has a non-saddle-point structure.
We observed the same situation in the previous section
in case of integral in $T$.

Let us exclude the integration in $T_0$ from the saddle-point approach
consideration. The set of other variables
$\{T_\alpha\}_{\alpha=1}^{r-1}$ has a zero solution at the saddle point.
So the stationarity solution for $u_{st}$ depends on $T_0$
only. As this non-saddle-point mode in the replica space is
chosen in an arbitrary manner, in fact we have constructed
$r$ identically instanton solutions in the replica space. This
results in the factor $r$ which allows us to produce the
operation $\lim\limits _{r\to 0}{\partial }/{\partial r}$
correctly.

The integration in $T_0$ must be treated in the
same way as the integration in $T$ in the previous section  and
yields the same result
\begin{equation}
\label{zak}
\mathop{\mbox{res}}\limits_{\varepsilon\to 0}\ln Z_{\nu}=
Const N^{Const}K_0^N/N!,\qquad N\to\infty.
\end{equation}
The calculation of $Const$ in this formula could not be produced
without an explicit calculation of the fluctuation integral.
But this is difficult and cumbersome problem that is not to be
solved here. Thus the asymptotic form (\ref{zak}) demonstrates
that the series investigated has a finite radius of
convergence.

In fact this section can be resumed as follows.
We have shown the saddle-point method yields the appropriate
result namely the circle of convergence for function $\ln G$
is determined by the same singularity as function $G$. The properties of this
singularity were calculated by saddle-point method. Other singularities for $\ln G$ could exist in
principle but these have non-saddle-point structure.

\section{Conclusions}
We have constructed the family of instantons in Kraichnan model
with "frozen" velocity field and found one of them in
explicit form.  Considering the asymptotic behaviour of the
renormalization constant at large order of perturbation
expansion we have demonstrated that the corresponding perturbation series
 has finite convergence radius. Furthermore, we have
 disproved the common statement that the behaviour of the
 series may be defined by quantity of diagrams at large order
of perturbation.

Our results can be used for an improvement of resummation
procedures constructed for the Kraichnan model.

The article was supported by RFBR (Grant No.08-02-00125a) 2008-2011.

\section{Appendix}

Let us consider the stationarity equations for $\x$,
$\x_0$, $t_0$ variables. The linear translation of fields
\begin{equation}
\label{trans}
\c_1(\tau_1)=\x_1+\x^{(1)}\frac{(\tau_1-t_1)}{T_1}+\bar{\c}_1(\tau_1),\quad
\c_2(\tau_2)=\x_0+\x^{(2)}\frac{(\tau_2-t_0)}{T_2}+\bar{\c}_2(\tau_2)
\end{equation}
shows explicitly the dependence of action $S$ on $\x_1$, $\x_2$.
The boundary conditions for new fields $\bar\c_l$ are assumed to be zero.
The differentiation of action $S$ in $\x_0$,$t_0$ produces cumbersome terms
due to the interaction part $S_u$ of the action. These terms
are of  the form
\begin{equation}
\label{t}
\frac{u}{2}\int d\tau_1 d\tau_1'c_1'(\tau_1')D'(\c_1(\tau_1')-
\c_1(\tau_1))c_1'(\tau_1)\cdot\frac{\tau_1-\tau_1'}{T_1},
\end{equation}
where $D'$ is a derivative of correlator $D$ on its argument (\ref{D(x)}).

Due to $D$ correlator is an even function, it's enough
to calculate in Exp. (\ref{t}) only the term corresponding to the
$\tau_1$ contribution. Substitution of particular solution $c_1'$
(\ref{sol}) yields
$$-\frac{u}{\nu^2 T_1}\int
d\tau_1'\frac{\pd c_1(\tau_1')}{\pd \tau_1'}\int d\tau_1
\tau_1 \frac{\pd D(c_1(\tau_1')-c_1(\tau_1))}{\pd
\tau_1}.$$

The inner integral can be calculated by parts.
Then the integral term is proportional to $\int
d\tau_1'[D_{11}\c_1'](\tau_1')$ and surface terms can
be calculated trivially.  The result turns out to
be calculated with the help of the following identities based on
the stationarity equations $$\int d\zeta \frac{\delta
S}{\delta \c_l'(\zeta)}=0\quad\Rightarrow\quad\int d\zeta
u[D_{ll}\c_l'](\zeta)+u[D_{12}(\c_1'+\c_2'-\c_l')](\zeta)=$$
$$=\int d\zeta(-2\nu c_l'(\zeta)+i\partial_\zeta c_l)=-i\nu x^{(l)},
\qquad l=1,2.$$
The last equality is written using the eqns. (\ref{sol}).

In this way the stationarity equation discussed
has the form simplified by $F=0$ condition:
$$
\frac{\delta S}{\delta t_0}=0\quad\Rightarrow \quad
\Bigg[\frac{x^{(2)}}{T_2}-\frac{x^{(1)}}{T_1}\Bigg]
\int\limits_{-x^{(1)}}^{x^{(2)}}D(z)dz=0,
$$

$$
\frac{\delta S}{\delta \x}=0\quad\Rightarrow
\quad \frac{iq\nu^2 }{u}=\int\limits_{0}^{x}D(z)dz.
$$
Besides we have the boundary conditions (\ref{bound1}) imposed by our
choice of particular solution $F=0$.
Let's note that the non-trivial instanton equation
with respect to $\x_0$ variable turn out to be an identity at $F=0$.
Indeed the substitution
of (\ref{sol}) into the action $S$ (\ref{S}) yields $S(\x)$ as the
function  in the unique $\x\equiv \x_2-\x_1$ space variable.
But the boundary conditions (\ref{bound1}) give an oportunity
to determine stationary value of $\x_0$.

\end{document}